\documentstyle[epsf]{mn}

\def \etal {{\rm et al.}}

\begin{document}

\title[Star formation in damped Ly$\alpha$ systems]
{Limits on the star formation rates of ${\mathbf z>2}$ damped Ly$\alpha$
systems from H${\mathbf \alpha}$ spectroscopy}

\author[A. J. Bunker {\rm et al. }]
 {Andrew~J.~Bunker,$^{1,2}$ Stephen~J.~Warren,$^{3}$ D.~L.~Clements,$^{4}$ \\
\vspace{-1.5mm}\\ {\LARGE Gerard~M.~Williger\,$^{5,6}$ and
Paul~C.~Hewett\,$^{7}$}\\ 
$^{1}$University of Oxford, Department of
Physics, Astrophysics, Keble Road, Oxford OX1 3RH\\ 
$^{2}$Present address: University of California at Berkeley, Department
of Astronomy, 601 Campbell Hall, Berkeley, CA~94720, USA\\ {\tt email:
bunker@bigz.Berkeley.EDU}\\ 
$^{3}$Imperial College, Department of Physics, Blackett Laboratory, 
Prince Consort Road, London SW7~2BZ\\
$^{4}$Department of Physics and Astronomy, University of Wales Cardiff,
P.O.\ Box 913, Cardiff, Wales CF23YB\\ 
$^{5}$MPIA, K\"onigstuhl 17, D-69117 Heidelberg, Germany\\
$^{6}$Present address: NASA Goddard Space Flight Center, Code 681, Greenbelt, 
Maryland 20771, USA\\ 
$^{7}$Institute of Astronomy, Madingley Road, Cambridge CB3~0HA\\ } 
\date{Accepted \\
Received }

\maketitle

\begin{abstract} We present the results of a long-slit $K$-band
spectroscopic search for H$\alpha$ emission from eight damped Ly$\alpha$
absorbers (DLAs) at $z>2$ with the goal of measuring the star-formation
rates in these systems. For each system we searched for {\it compact}
sources of H$\alpha$ emission within a solid angle $11\times
2.5$~arcsec$^{2}$ ($44\times 10~h^{-2}$~kpc$^{2}$, for $q_{0}=0.5$,
$H_{0}=100~h~{\rm km~s}^{-1}~{\rm Mpc}^{-1}$). No H$\alpha$ emission was
detected above $3\sigma$ limits in the range $(6.5 - 16)\times
10^{-20}$~W~m$^{-2}$, equivalent to star formation rates of $(5.6 -
18)~h^{-2}~{\rm M_{\odot}~yr}^{-1}$, for a standard IMF, assuming the
lines are spectrally unresolved ($<650$~km~s$^{-1}$ FWHM).  We compare
these results against the predictions of the models of Pei \& Fall
(1995) of the global history of star formation, under two different
simplifying hypotheses: i.) the space density of DLAs at $z=2.3$ is
equal to the space density of spiral galaxies today (implying DLA disks
were larger in the past, the `large-disk' hypothesis); ii.) the sizes
of DLAs at $z=2.3$ were the same as the gas sizes of spiral galaxies
today (implying DLA disks were more common in the past, the
`hierarchical' hypothesis). Compared to the previous most sensitive
spectroscopic search our sample is twice as large, our limits are a
factor greater than two deeper, and the solid angle surveyed is over
three times as great. Despite this our results are not in conflict with
either the large-disk hypothesis, because of the limited solid angle
covered by the slit, or the hierarchical hypothesis, because of the
limited sensitivity.
 
\end{abstract} 

\section{Introduction}
\label{sec:intro}
 
The history of galaxies -- when and how they formed, and how they have
evolved -- is a topic of enormous current interest. The traditional
observational approach has been through deep imaging to define a
flux-limited sample of galaxies. From the measurement of brightnesses
and redshifts the luminosity function at different epochs can be
constructed. There has been rapid progress in this field over the past
three years, beginning with the measurement by Lilly \etal\ (1996) of the
global star formation rate in galaxies out to $z=1$, continuing with
the remarkable success of Steidel \etal\ (1996) in securing redshifts
for dozens of galaxies at $z\approx 3$, and culminating in the plot of
Madau \etal\ (1996) depicting for the first time a measurement of the
history of global star formation in galaxies from today back to $z=4$.

The measurement of quasar absorption lines allows an independent
approach to studying the history of galaxies. The highest hydrogen
column density absorbers seen in the spectra of background QSOs have
$N({\rm H{\scriptstyle~I}})\ga 2\times 10^{24}$~m$^{-2}$. These damped
Ly$\alpha$ absorption systems (DLAs, Wolfe \etal\ 1986) contain most
of the neutral gas in the Universe (Lanzetta, Wolfe \& Turnshek
1995). From the redshift distribution of the damped systems, and the
measured column densities, the evolution in the co-moving density of
neutral gas $\Omega_g$ can be measured (e.g., Lanzetta \etal\ 1991)
provided the consequences of dust obscuration are accounted for. Then
the analysis of the variation of $\Omega_g$ with redshift allows the
measurement of the history of star formation in the Universe (Pei \&
Fall 1995).

The latter, spectroscopic, approach to the history of star formation in
galaxies unfortunately tells us nothing about how galaxies are
assembled, and the nature of DLAs is currently debated. One school of
thought has DLAs being the (large) progenitors of massive spiral disks
(e.g., Wolfe \etal\ 1986; Lanzetta \etal\ 1991). Evidence in support of
this interpretation includes the measurement of line velocity profiles
of low-ionization species consistent with those expected from
lines of sight intercepting thick gaseous disks of circular velocity
$\sim 200~{\rm km~s}^{-1}$ (Prochaska \& Wolfe 1997). In this case,
however, it is difficult to reconcile the low metallicities of
high-redshift DLAs (average metallicity 1/13 of solar at $z\approx 2.5$,
Pettini \etal\ 1997) with the metallicities of stars in spiral disks
today. The proposed cold rotationally-supported
disks with these large circular speeds run counter to the predictions of
cosmogonical theories currently in vogue. However Haehnelt, Steinmetz \&
Rauch (1998), using CDM hydrodynamic simulations, have shown that
irregular (small) proto-galactic clumps can give rise to absorption
profiles similar to those measured by Prochaska \& Wolfe. In addition,
M\o ller \& Warren (1998) have recently shown that the impact
parameters of the few detected galaxy counterparts of high-redshift DLAs
are small (in the context of this debate) and that the space density of
the DLAs at high-redshift is probably much higher than the space density
of spiral galaxies today. In this paper we refer to these different
pictures as the `large-disk' and `hierarchical' hypotheses.
 
There have been extensive searches for Ly$\alpha~\lambda~121.6$~nm
emission associated with star formation in DLAs but with limited
success (e.g., Smith \etal\ 1989; Hunstead, Pettini \& Fletcher 1990;
Wolfe \etal\ 1992; M\o ller \& Warren 1993; Lowenthal \etal\
1995). This is generally thought to be due to resonant scattering
greatly extending the path length of Ly$\alpha$ photons escaping
through a cloud of neutral hydrogen so that even very small quantities
of dust can extinguish the line (Charlot \& Fall 1991). Because the
effective extinction can be extremely large this has the consequence
that non-detections do not provide any useful information on the star
formation rates in the DLAs. The H$\alpha~\lambda$~656.3~nm line,
although intrinsically weaker by a factor $\approx$~10, lies at a longer
wavelength where the extinction is smaller, and is not resonantly
scattered. In consequence a search for H$\alpha$ emission from DLAs may
be more efficient than a search for Ly$\alpha$.
 
In this paper, we present the results of a spectroscopic survey for
H$\alpha$ emission from eight damped absorption
systems at $2.0<z<2.6$, along the line-of-sight to six high-redshift
quasars. At these redshifts H$\alpha$ appears in the $K$-window in the
near-infrared. The results are relevant to the debate on the nature of
the DLAs, for, as shown in \S5, if the DLAs are
the counterparts of today's spiral galaxies the associated H$\alpha$
emission is within the reach of current instrumentation. The
calculation presented there uses the analysis by Pei \& Fall (1995) of
the global star formation rate. The average star formation rate in
each DLA depends then on their space density, so that high measured
rates of star formation would provide support for the view that the
DLAs are massive galaxies already in place at high-redshift. A low measured
star formation rate on the other hand would be in agreement with the
hierarchical picture.
 
The layout for the remainder of the paper is as follows. In \S2 we
describe the near-infrared spectroscopic observations and in \S3 the
data reduction. The results are used to constrain the star formation
rates in \S4. The reader who is not interested in the technicalities of
the survey could proceed directly to \S5, where we compare these results
against the predictions of the large-disk and hierarchical hypotheses.
In \S\ref{sec:conclusions} we consider the implications for strategies
for detecting emission from DLAs and summarise our conclusions.
Appendix~\ref{app:qsospec} gives the results from the spectroscopy of
the background quasars, and Appendix~\ref{sec:phl957c1spec} describes
the near-infrared properties of a companion galaxy to the $z=2.31$ DLA
PHL957 (Lowenthal \etal\ 1995). We assume a cosmology with ${\rm
H_{0}}=100~h$~km~s$^{-1}$~Mpc$^{-1}$, $\Lambda_{0}=0$ and $q_{0}=0.5$
unless otherwise stated.

\section{Observations}
\label{sec:obs}

The spectra were obtained over the nights 1995 October 17 to 19 UT at
the 3.8-m United Kingdom Infrared Telescope (UKIRT) with the CGS\,4
instrument. Each night was partly cloudy. The journal of observations is
provided in Table~\ref{tab:obs}. This lists target name, target
coordinates as measured off the Digitized Sky Survey, date of
observation, total integration time, and slit position angle (PA)
measured east of north. The redshifts of the absorbers are provided in
Table~\ref{tab:dlalimits}.  The detector is a $256^{2}$ InSb array. The
instrument was equipped with the 150~mm focal length `short' camera, the
75~lines~mm$^{-1}$ grating, and the 2~pixel slit. In this configuration
the pixel size corresponds to 1.25~arcsec in the spatial direction and
the resolving power is $\frac{\lambda}{\delta\lambda}\approx 450$
(confirmed by measuring the spectral width of arc and sky lines).
The seeing varied within the range 0.7--1.2~arcsec, and was therefore
much less than the slit width of 2.5~arcsec.

The targets were acquired by peaking-up on a nearby bright Carlsberg
Meridian Circle star, zeroing the telescope coordinates, and then moving
to the QSO position. The combined accuracy associated with the
acquisition and astrometry is estimated to be $\sim0.5$~arcsec. With the
exception of one target the central wavelength was set to $2.17~\mu$m,
which placed the wavelength range $1.86~\mu{\rm
m}\la\lambda\la2.52~\mu{\rm m}$ on the array and captured the entire
$K$-window.  For one of the two PAs used in spectroscopy of PHL957 the
grating angle was set to provide coverage of the wavelength interval
$1.58~\mu{\rm m}\la\lambda\la2.24~\mu{\rm m}$, which includes the lines
H$\alpha$, [O{\scriptsize~III}]~$\lambda\lambda$~495.9,500.7~nm and
H$\beta$~486.1~nm at the redshift of the DLA ($z=2.31$). For this target
the slit was oriented to include a nearby companion galaxy at the same
redshift (Lowenthal \etal\ 1991). The spectrum of the companion galaxy
is presented in Appendix~\ref{sec:phl957c1spec}.

The observing procedure followed was to nod along the slit between two
positions 11--15~arcsec apart, keeping the source on the chip,
and following the sequence {\em ABBAABB...} At each position {\em A} or
{\em B} the detector was stepped mechanically over four half-pixel
increments. This enables over-sampled spectra to be obtained and
eliminates gaps in the spectrum due to isolated bad pixels.  At each
step two 15~s integrations were co-added, and then the summed
integrations at the four steps were interleaved to produce a single
`observation' comprising 2~min of integration. The individual
integration time of 15~s was set so as to avoid saturation of the
strongest OH lines while ensuring that at all wavelengths the sky
Poisson noise dominated the read out noise, which was approximately
$ 15~e^{-}$. The dark current of CGS\,4 is negligible,
$<1~e^{-}~{\rm s}^{-1}$.

\begin{table}
\centering
\begin{tabular}{l|c|c|c|c|c}
QSO & Coords & Date & Int.\
& Slit & 
 \\
& (B~1950.0) & (UT) & time & PA \\
& & dd-mm-yr & /min & /$^{\circ}$ \\
\hline\hline
& & & & \\
PHL957 &
01$^{\rm h}$00$^{\rm m}$33\fs39 & 17-10-95 & 56 & $-75.9$ \\
& $+$13\degr00\arcmin10\farcs6 &
19-10-95 & 68 & $-60.5$ \\
& & & & \\
0458$-$020 &
04$^{\rm h}$58$^{\rm m}$41\fs28 & 17-10-95 & 74 & $-157.9$ \\
 & $-$02\degr03\arcmin34\farcs5 & & & \\
& & & & \\
0528$-$250 &
05$^{\rm h}$28$^{\rm m}$05\fs21 & 18-10-95 & 32 & $-34.0$ \\
& $-$25\degr05\arcmin44\farcs7 & 19-10-95 &
24 & $-66.5$ \\
& & & & \\
2343$+$125 & 23$^{\rm h}$43$^{\rm m}$55\fs42 & 17-10-95 & 96 & $-35.0$ \\
 & $+$12\degr32\arcmin19\farcs8 & & & \\
& & & & \\
UM184 &
23$^{\rm h}$48$^{\rm m}$24\fs02 & 17-10-95 & 70 & $-166.4$ \\
& $-$01\degr08\arcmin51\farcs3 &
18-10-95 & 56 & $-13.6$ \\
& & & & \\
UM196 & 23$^{\rm h}$59$^{\rm m}$16\fs24 & 18-10-95 & 64 &
$-71.4$ \\
& $-$02\degr16\arcmin22\farcs6 & & & \\
& & & & \\
\end{tabular}
\caption{Journal of spectroscopic observations.}
\label{tab:obs}
\end{table}

\section{Data Reduction}
\label{sec:reduction}

The first stage in the data reduction was to divide each sub-frame by a
normalised flat-field frame. The four sub-frames, stepped at half-pixel
intervals, were then interleaved, with logged bad pixels ignored. This
produced a frame at each nod position with 514 sample points along the
dispersion axis, from the original 256~pixels. By tracing the position
of arc lines across the argon arc reference frames a two-dimensional
(2D) dispersion solution, rms 0.2~nm, was computed and the frames were
transformed on to a linear wavelength scale, with dispersion 1.32~nm per
pixel. To compensate for non-uniformities in the slit profile an
illumination correction was applied, determined by collapsing the sky
along the dispersion axis.

First-order sky subtraction was achieved by forming the difference image
of the adjacent frames at positions $A$ and $B$. The $A-B$ pairs
for each QSO and each slit PA were then summed separately using a
sigma-clipping algorithm to eliminate discrepant data values, due, e.g.,
to cosmic rays. Residual sky counts were then removed by fitting a
low-order polynomial to each column in the combined frame. The frames at
this stage therefore contain a positive spectrum at position $A$ and
an independent negative spectrum at position $B$, so that a
combination was formed by subtracting from the original a shifted
version of the frame, with the two spectra registered.

The spectra were corrected for atmospheric absorption and flux
calibrated using observations of A and F spectral-type standard
stars. Stellar absorption features in the spectra of the standard stars
were firstly interpolated across. The standard-star spectra were then
divided into a black-body curve of the appropriate temperature, scaled
to the flux level corresponding to the known $K$ magnitude. The
resulting curves were used to calibrate the spectra. The calibration
curves from night to night agreed to within 5~per~cent, so the standards do
not appear to have been affected by clouds. It was then possible to
check the effect of clouds on the quasars by comparison of the $K$-band
magnitudes measured from the calibrated spectra with the results from
our own calibrated $K$-band images obtained in a complementary programme
(Bunker \etal\ 1995; Bunker 1996). With one exception, UM196, we found
good agreement. For the case of UM196 the spectroscopic magnitude was
$0.8^{m}$ fainter than that measured in the image. This was assumed to
be due to cloud and was therefore corrected for.
 
\begin{figure}
\epsfxsize=\columnwidth \epsfbox{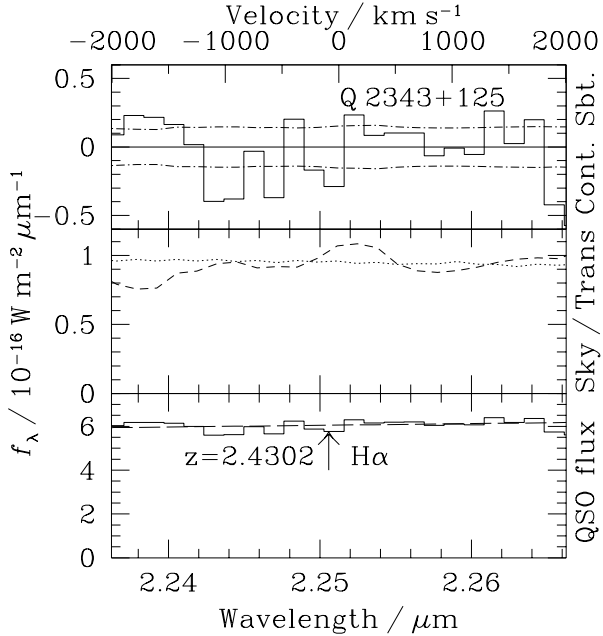}
\caption{Example of the data showing the spectrum of the QSO
2343+125 ($z_{\rm abs}=2.4302$) over the velocity range
2000~km~s$^{-1}$ either side of the predicted wavelength of H$\alpha$
at the redshift of the damped Ly$\alpha$ system seen in the QSO
spectrum. The {\it bottom} panel shows the QSO spectrum (solid line)
with the continuum fit (long-dash line). The {\it middle} panel shows
the fractional atmospheric transmission (dotted line) and the sky
spectrum scaled down by a factor of 1000 (short-dash line). The {\it
top} panel shows the continuum-subtracted spectrum (solid line) and the
$\pm1\sigma$ noise per pixel (dot-dash). }
\label{fig:dampedIRspec}
\end{figure}

In order to search for line emission from the damped systems it was
necessary to subtract accurately the flux from the background QSO. This
was relatively straight-forward as the continua are smoothly-varying
with wavelength and fairly flat in $f_{\nu}$. In no case was there any
confusion with lines from the QSO.  A low-order function, usually a
first-order cubic spline, was fitted to each row occupied by the QSO in
the 2D spectrum around the predicted H$\alpha$ wavelength, excluding the
region within $\sim 1000$~km~s$^{-1}$ of the expected line wavelength.
An example of the results is shown in Fig.~\ref{fig:dampedIRspec} which
plots the portion of the spectrum of the QSO 2343+125 around H$\alpha$,
together with the continuum fit, and the continuum-subtracted spectrum.
The noise was computed on the basis of Poisson statistics and is also
shown in the plot. Because the quasars are $\ga 20$ times fainter than
the sky in $K$ the contribution of the shot noise from the QSO spectrum
is minimal. We also computed the standard deviation of the counts in the
sky at each wavelength, up each column, as a check on the Poisson
estimate of the noise, and found good agreement. The sensitivities
reached (Tables~1 \& 2) are similar to those quoted in the CGS\,4
manual, which for a 1~pixel-wide slit (1.25~arcsec) are a $3\sigma$
detection of unresolved line emission at 2.2~$\mu$m in 30 min of
$8\times 10^{-20}$~W~m$^{-2}$.

\begin{table*}
\begin{center}
\begin{tabular}{|l|c|c|c|c|c|c|c|c|}
\hline
QSO & $z_{\rm abs}$ & $n({\rm H{\scriptstyle~I}})$ & Refs. &
$f_{\rm H\alpha}$~($3\sigma$) & \multicolumn{2}{c}{$q_{0}=0.5$} &
\multicolumn{2}{c}{$q_{0}=0$} \\
& & $10^{24}$~m$^{-2}$ & & $10^{-20}$~W~m$^{-2}$ & $L_{\rm H\alpha}$~($3\sigma$)&
SFR~($3\sigma$) & $L_{\rm H\alpha}$~($3\sigma$) &
SFR~($3\sigma$)\\
& & & & & $10^{34}~h^{-2}$~W &
$h^{-2}~{\rm M_{\odot}~yr}^{-1}$ & $10^{34}~h^{-2}$~W &
$h^{-2}~{\rm M_{\odot}~yr}^{-1}$ \\
\hline\hline
PHL957 (both) & 2.3091 & 25~$\pm$~3 & 1, 2, 3 &  6.52 & 6.24 & 5.57 & 17.4 & 15.5\\ 
PHL957 (PA$=-75.9^{\circ}$) & & & & 9.70 & 9.28 & 8.28   &    25.8  &   23.1\\ 
PHL957 (PA$=-60.5^{\circ}$) & & & & 8.80 & 8.42 & 7.52   &    23.4  &   21.0\\ 
0458$-$020 & 2.03950  &  45~$\pm$~10 & 1, 3, 4 & 9.03 & 6.53 & 5.83 & 16.5 & 14.7 \\
0528$-$250 (both) & 2.14030  &  5 & 5 & 10.0 & 8.09 & 7.23   &     21.2  &   18.9\\ 
0528$-$250 (PA$=-34.0^{\circ}$) & & & & 13.3 & 10.7  & 9.57  &    28.1   &  25.1\\ 
0528$-$250 (PA$=-66.5^{\circ}$) & & & & 15.3 & 12.4 & 11.0   &    32.4   &  28.9\\ 
2343$+$125 & 2.4302 & 3 & 6 &  7.02 & 7.53 &    6.73   &    21.9  &   19.6\\ 
UM184a (both) & 2.6145 & 18~$\pm$~3 & 3, 5, 7 &   10.5 & 13.3 &    11.9 & 41.3  &   36.8\\
UM184a (PA$=-166.4^{\circ}$)  & & & &    14.1 & 17.9 &    16.0 &   55.3  &   49.4 \\
UM184a (PA$=-13.6^{\circ}$)  & & & &    15.8 & 20.0 &    17.8  &   61.9  &   55.2 \\
UM184b (both) & 2.4261 & 3.0~$\pm$~0.5 & 3, 7 &  7.53 & 8.05 &    7.18 & 23.4  &   20.8\\
UM184b (PA$=-166.4^{\circ}$)  & & & & 10.1 & 10.8 & 9.64  &  31.3  &   28.0\\ 
UM184b (PA$=-13.6^{\circ}$)  & & & & 11.3 & 12.1 & 10.8   &  35.0  &   31.3\\ 
UM196a & 2.1537 & 2.0 & 4 & 10.0 & 8.21 &   7.33    &   21.6  &   19.3\\ 
UM196b & 2.0951 & 4.5 & 4 &  10.5 & 8.10 &    7.23   &    20.9  &   18.6\\ 
\end{tabular}
\end{center}
References:
(1)~Meyer \& Roth (1990);
(2)~Pettini, Boksenberg \& Hunstead(1990);
(3)~Pettini \etal\ (1994);
(4)~Wolfe \etal\ (1993);
(5)~Sargent, Steidel \& Boksenberg (1989);
(6)~Sargent, Boksenberg \& Steidel (1988); 
(7)~Turnshek \etal\ (1989).
\caption{The $3\sigma$ upper limits for H$\alpha$ emission from the eight
damped Lyman-$\alpha$ systems. This assumes the lines are unresolved
($<~650$~km~s$^{-1}$ FWHM), that all the star formation from the DLA
is concentrated in a region smaller than $1.25\times2.5$~arcsec$^2$, and
that this region of star formation falls somewhere in the
$11\times2.5$~arcsec$^2$ 
solid angle searched. Where the data were taken at two different PAs, we
quote the sensitivity in the region of overlap (centred on the QSO), as
well as for the individual PAs.}
\label{tab:dlalimits}
\end{table*}

\section{Results: Upper limits on line emission from the absorbers}
\label{sec:results}

The choice of aperture size for extraction of the spectra depends on
the expected spatial distribution of the regions of star formation in
the DLAs. It should be noted that under the large-disk hypothesis the
DLAs have large radii, typically $R\approx 30~h^{-1}$ kpc at $z\approx
2.3$ for $q_{0}=0.5$, or $\approx 8$~arcsec. If the H$\alpha$ emission
is spread approximately uniformly over the disk then a large aperture
matching the disk size reaches the deepest flux limit. If on the other
hand most of the star formation is concentrated at one point the
deepest flux limit is achieved by searching for emission lines in
several parallel apertures, along the slit, each of width equal to the
expected size of the region of star formation. The strategy we adopted
was to use nine spatially contiguous apertures of size ${\mathrm
1~pixel}=1.25$~arcsec, one centred on the quasar, and four on either
side. (The 2D long-slit spectra were also carefully inspected at
larger offsets along the slit to ensure that no line emission
intersected by the slit at greater impact parameters was overlooked.)
The quoted sensitivities for the search are valid provided: i.) most
of the star formation from any DLA is compact and concentrated in a
solid angle smaller than $1.25\times2.5$~arcsec$^2$, equivalent to
$\approx 5.0\times10.0~h^{-2}$ kpc$^2$, ii.)  the region of star
formation falls within the $11\times2.5$~arcsec$^2$ solid angle
searched. The first assumption is certainly true of one DLA, the
system at $z=2.81$ towards the quasar 0528$-$250 (M\o ller \& Warren
1998), for which the half light angular radius is
$r_{0.5}=0.13\pm0.06$~arcsec. The optical angular sizes of the Lyman-break
galaxies (Steidel \etal\ 1996) are similarly small (Giavalisco \etal\
1996), and this rest-UV morphology traces the star forming regions.
In \S5 we quantify the fraction of DLAs for which the region of star
formation would fall outside the solid angle searched.

We have made a search for line emission in the extracted one-dimensional
(1D) spectra (extraction width $1~{\rm pixel}=1.25$~arcsec), near the expected
wavelength of H$\alpha$, by integrating the counts in each
continuum-subtracted spectrum over 4 pixels spectrally (i.e., 5nm, just
larger than the resolution of FWHM=650~km~s$^{-1}$) at all wavelengths
within $\pm$2000~km~s$^{-1}$ of the predicted wavelength. No lines were
detected at $>3\sigma$ significance above the quasar continuum. For
our $1.25~{\rm arcsec}\times 2.5~{\rm arcsec}\times 5$~nm aperture, the
$3\sigma$ limits lie in the range $(6.5 - 16)\times
10^{-20}$~W~m$^{-2}$ and are recorded in Table~\ref{tab:dlalimits}.
Note that the resolution of the spectra is sufficient to separate
H$\alpha$ from the [N{\scriptsize~II}] doublet so no correction to the
H$\alpha$ flux to account for blending is necessary. A similar search
was made for [O{\scriptsize~III}]~$\lambda$~500.7~nm and H$\beta$ in the
$H$-band spectrum of PHL957. No line emission was detected above the
$3\sigma$ limit of $5.3\times 10^{-19}~{\rm W~m}^{-2}$.

Also recorded in Table~\ref{tab:dlalimits} are the H$\alpha$ line
luminosities corresponding to the line-flux limits, and the inferred
limits to the star formation rates in these systems, based on the
prescription of Kennicutt (1983), where a star formation rate (SFR) of
$1~{\rm M_{\odot}~yr}^{-1}$ generates a line luminosity in H$\alpha$ of
$11.2\times 10^{33}$~W. The limits to the SFRs lie in the range $(5.6 -
18)~h^{-2}~{\rm M_{\odot}~yr^{-1}}$. The conversion between H$\alpha$
line luminosity and SFR is uncertain and depends on the assumed initial
mass function (IMF). Kennicutt adopted an IMF similar to that of
Salpeter (1955). For comparison, from the models of Bruzual \& Charlot
(1993) a Scalo (1986) IMF with stars in the mass range $0.1~{\rm
M_{\odot}}<{\rm M_{*}}<125~{\rm M_{\odot}}$ yields an H$\alpha$ line
luminosity of $L({\rm H}\alpha)=9.4\times 10^{33}$~W for a star
formation rate of $1~{\rm M_{\odot}~yr}^{-1}$ (e.g., Gallego \etal\
1995).

To date there have been no published successful detections of emission
lines in the near-infrared from high-redshift DLAs. A detection of
[O{\scriptsize II}]~$\lambda$~372.7~nm and H$\beta$ by Elston \etal\
(1991) from DLA 1215+333 was not confirmed in subsequent observations
(J.~Lowenthal, {\em private communication}). The only comparable limits
are those of Hu \etal\ (1993) who observed three DLAs and recorded
$3\sigma$ upper limits to the H$\alpha$ flux of $\approx
3\times10^{-19}$~W~m$^{-2}$. The solid angle searched was either
$3\times 3$~arcsec$^{2}$ or $3\times 1$~arcsec$^{2}$ (E.~Hu, {\em
private communication}). The survey presented here, therefore, is more
than twice as large, reaches flux limits over twice as deep, and covers
a solid angle for each DLA over three times greater.

For comparison, the characteristic star formation rate of the Lyman
break population at $z\sim 3 - 4$ is $16~h^{-2}~{\rm
M_{\odot}~yr}^{-1}$ (Steidel \etal\ 1999) for $q_{0}=0.5$. This is
determined from the rest-UV continuum around $\lambda_{\rm rest}\approx
1500$~\AA, assuming a Salpeter IMF and correcting for a mean
extinction of $E(B-V)\approx 0.15$ using the Calzetti (1997) reddening
law. Therefore, our $3\sigma$ limiting fluxes, equivalent to star
formation rates of $(5.6 - 18)~h^{-2}~{\rm M_{\odot}~yr^{-1}}$ at
$z\approx 2.3$, probe SFRs which are sub-$L^{*}$ for the Lyman break
galaxies. Indeed, our deepest flux limit is sensitive to
Lyman-break-type galaxies as faint as $0.35~L^{*}$.  However, the
relationship between these Lyman break galaxies and DLAs is unclear from
the few examples of $z>2$ DLAs detected in both optical imaging and
spectroscopy. The $z\approx 2.8$ DLA 0528-250 (M\o ller \& Warren 1993;
Warren \& M\o ller 1996; M\o ller \& Warren 1998) has a star formation
rate of only $\sim 1~h^{-2}~{\rm M_{\odot}~yr^{-1}}$ (which would fall
below our H$\alpha$ survey limit) based on the relatively unextinguished
Ly$\alpha$ emission with a large rest-frame equivalent width of
$W_{0}\approx5$~nm. This is quite unlike most Lyman break galaxies,
which typically exhibit much higher star formation rates ($\sim
10~h^{-2}~{\rm M_{\odot}~yr^{-1}}$) and much lower Ly$\alpha$ equivalent
widths ($W_{0}\approx0.3-2$; Steidel \etal\ 1996). However, the $z=3.15$
absorber towards 2231+131 is more typical of the Lyman break population,
and was indeed selected as such through broad-band imaging (Steidel,
Pettini \& Hamilton 1995).  Spectroscopy showed Ly$\alpha$ in emission
with a rest-frame equivalent width of $W=2.3$~nm (Djorgovski \etal\
1996), and the rest-frame UV continuum is around $L^{*}_{1500}$ for the
Lyman-break galaxies, equating to star formation rates above our
$3\sigma$ H$\alpha$ survey limit if the reddening is $E(B-V)\ga 0.1$.
The impact parameters in both these cases are small: 1.2~arcsec for
0528-250 ($4.3~h^{-1}$ projected); 2.3~arcsec for 2231+131
($8.3~h^{-1}$ projected).

\section{Discussion: comparison of results against predictions of the
large-disk and hierarchical hypotheses.}
\label{sec:discussion}

In this section we compare the measured upper limits on the star
formation rates in our sample against theoretical predictions.
We start with the analysis by Pei \& Fall (1995) of the observed rate
of decline of the cosmic density of neutral gas $\Omega_g(obs)$
measured from surveys for DLAs. At any redshift $\Omega_g(obs)$ will be
an underestimate of the true cosmic density $\Omega_g(true)$ because
quasars lying behind DLAs will suffer extinction, and may therefore
drop out of the samples of bright quasars used to find the DLAs. Pei \&
Fall correct for this bias, accounting in a self-consistent manner for
the increasing obscuration as the gas is consumed and polluted as star
formation progresses. In this way they determine the evolution of
$\Omega_g(true)$, and so the SFR per unit volume: 
\begin{equation}
\dot{\rho}(z)=-{\rm
3H_{\circ}^2}\dot{\Omega}_{g}(true)/(8\pi G) . 
\end{equation}

The analysis of Pei \& Fall is global and gives no information on the
SFRs in individual galaxies. We adopt their predictions for the
volume-average SFR, $\dot{\rho}(z)$, and then need to make assumptions
about how many DLAs there are at the redshifts of interest, and how
$\dot{\rho}(z)$ is distributed between them. The methodology for the
comparison is set out in the next sub-section. We adopt two different
assumptions for the space density and sizes of DLAs, the `large-disk'
hypothesis and the `hierarchical' hypothesis, explained below.

\subsection{Methodology}

We begin with the parametrization of the properties of spiral galaxies
locally used by Lanzetta \etal\ (1991) in their analysis of the
statistics of DLAs, i.e., we adopt a $B$-band galaxy luminosity function
of Schechter form, with space density normalisation $\Phi^*$, power
index $s$, and spiral fraction $f_s$, a power-law (Holmberg) relation
between radius and luminosity $R/R*=(L/L*)^t$, and a ratio between gas
radius and optical (Holmberg) radius $\xi$, independent of
luminosity. The standard values of the various parameters are
$\Phi^*=1.2\times10^{-2}~h^3$~Mpc$^{-3}$, $s=1.25$, $f_s=0.7$, $t=0.4$,
$R*=11.5~h^{-1}$~kpc, and $\xi=1.5$ (Lanzetta \etal\ 1991).  $L*$ itself
is not needed in the analysis below.

At zero redshift we distribute the SFR density $\dot{\rho}(0)$ over the
galaxies by assuming the SFR in an individual galaxy is proportional to
the galaxy luminosity. The global star formation rate, $\dot{\rho}(z)$,
predicted by Pei \& Fall increases with redshift, peaking near
$z=2$. The SFRs in the DLAs will depend then on $\dot{\rho}(z)$ and on
the evolution in space density of the population. Now the average
incidence of DLAs per unit redshift, $dn/dz$, along a 1D sight-line to a quasar
is proportional to the product of the space density and the gas cross
section of the absorbers. Lanzetta \etal\ found that the incidence of
DLA absorbers at $z\approx 2.5$ was considerably higher than expected,
by factors of $F\approx 3.8$ for $q_{0}=0$ and $F\approx 7.1$ for
$q_{0}=0.5$, on the basis of no evolution in galaxy cross section or
space density. The mean redshift of our sample, $z=2.3$, is close to
that of the Lanzetta \etal\ sample. To use the information on the line
density we consider two different hypotheses. With the `large-disk'
hypothesis (LD) we assume that the space density of DLAs at high
redshift is equal to the space density of spiral galaxies locally. For
consistency with the line density of DLAs we suppose under this
hypothesis that all DLAs in the past were larger in radius by the factor
$\sqrt F$. In this case the SFR in each DLA was larger in the past by a
factor $\dot{\rho}(z)/\dot{\rho}(0)$.

An alternative hypothesis is that DLAs at high redshift had the same
distribution of gas sizes as spiral galaxies locally, and that the
space density was greater in the past by a factor $F$. We call this the
`hierarchical' hypothesis (H). In this case there are $F$ DLAs at
high redshift for every local spiral galaxy, each of the same size as
the local spiral, and the ratio of the SFR in each high-redshift DLA to
the SFR in the nearby spiral is $\dot{\rho}(z)/(F\dot{\rho}(0))$. Note
that this hierarchical hypothesis is merely a single specific
formulation of the hierarchical picture, and is motivated only by its
simplicity. It is not a generic representation of theories of
hierarchical galaxy formation. The two hypotheses LD and H, and two
cosmologies $q_{0}=0$ and $q_{0}=0.5$, give four models which we call
LD0, LD5, H0, and H5.

With these assumptions, the distribution functions of radius and SFR at
high redshift are fully specified for the four models. This allows a
direct comparison against our observational results.

\subsection{The mean SFR in a sample of DLAs at $<z>=2.3$}

Samples of DLAs are selected from spectra, so the relative numbers of
DLAs of different sizes are weighted by cross section. In this
sub-section we compute the mean SFR in such a sample of DLAs. This
provides a convenient measure to compare against our upper limits.
However the calculation takes no account of the possibility that the
region of star formation falls off the slit. In the next sub-section
we quantify the incompleteness due to the limited solid angle covered
by the slit.

The total luminosity density of spiral galaxies integrated over the
Schechter function is ${\cal L}=f_s \Phi^* L^* \Gamma(2-s)$ where
$\Gamma$ denotes the gamma function. For the large-disk
hypothesis the star formation rate in the DLA counterpart of a spiral
galaxy of present-day luminosity $L(0)$ is given by
\begin{equation}
{\mathrm{SFR}}(z)({\mathrm{LD}})=\frac{L(0)\dot{\rho}(z)}{f_s \Phi^*
L^*(0) \Gamma(2-s)} .
\end{equation}
 The average SFR in a sample of DLAs follows from
computing the gas-cross-section weighted average present-day luminosity
of a spiral galaxy, which is given by
\begin{equation}
\bar{L}(0)=\frac{\Gamma(2+2t-s)L^*(0)}{\Gamma(1+2t-s)} .
\end{equation}
 This gives
the following relation for the expected average SFR in a sample of DLAs
\begin{equation}
{\mathrm{\overline{SFR}}_{DLA}(LD)}=\frac{{\rm
-3{\mathrm H_{0}^2}}\dot{\Omega}_{g}(true)\Gamma(2+2t-s)}{8\pi
Gf_s\Phi^*\Gamma(2-s)\Gamma(1+2t-s)} .
\end{equation}
Inserting the standard values of the
various parameters we obtain finally
\begin{equation}
{\mathrm{\overline{SFR}}_{DLA}(LD)}=-1.48\times10^4~\dot{\Omega}_{g}(true)\hspace{1cm}h^{-1}~{\rm
M_{\odot}~yr}^{-1} ,
\end{equation}
where $\dot{\Omega}_{g}(true)$ is in units of Gyr$^{-1}$ as provided
by Pei and Fall (1995).

For the hierarchical hypothesis (H) we have
\begin{equation}
{\mathrm{\overline{SFR}}_{DLA}(H)=\mathrm{\overline{SFR}}_{DLA}(LD)}/F .
\end{equation}

\begin{figure}
\epsfxsize=\columnwidth \epsfbox{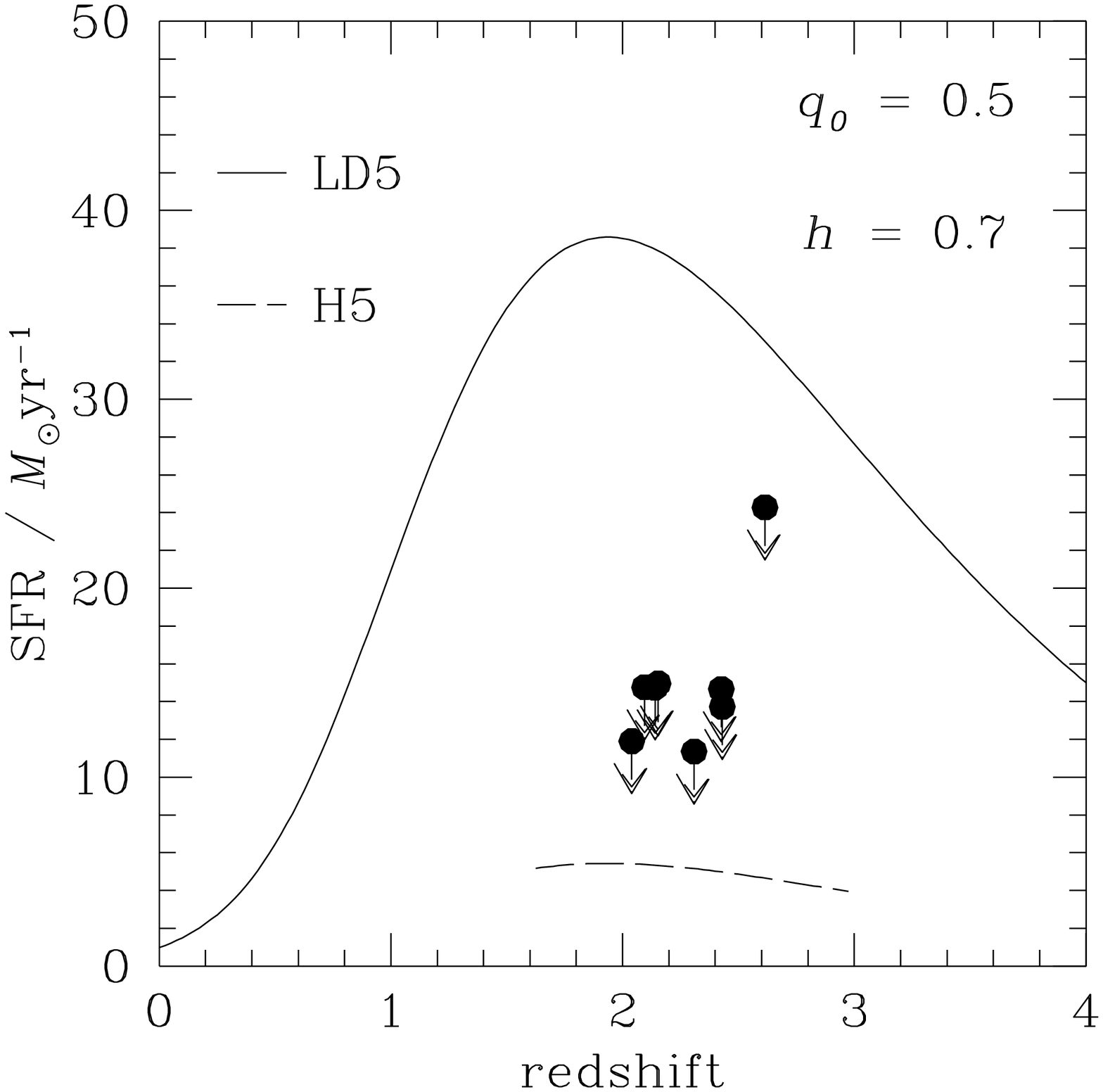}
\epsfxsize=\columnwidth \epsfbox{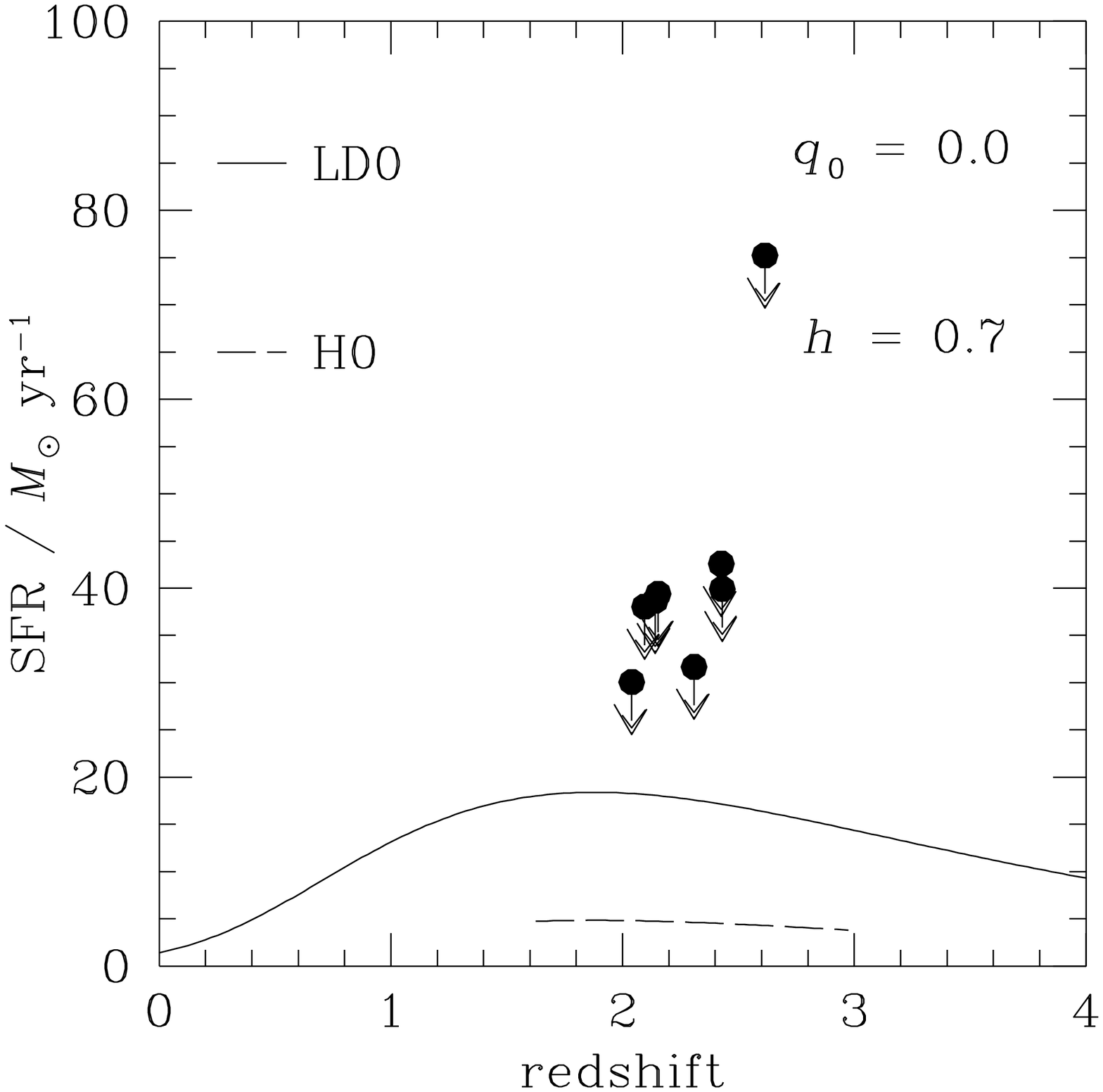}
\caption{Comparison of observational limits to the SFRs in the sample of
DLAs against theoretical predictions for $q_{0}=0.5$ (upper plot) and
$q_{0}=0.0$ (lower plot).  The measured 3$\sigma$ upper limits for the
1.25~arcsec~$\times$~2.5~arcsec~$\times$~5nm extraction aperture are
plotted as arrows. The solid curves are the predicted sample-averaged
(i.e., cross-section weighted) SFRs for the closed-box models of Pei \&
Fall (1995) for $\Omega_{g\infty}=4\times10^{-3}~h^{-1}$, and for the
large-disk hypothesis (LD). The dashed curves are the predictions under
the hierarchical hypothesis (H) and are valid only over the limited
redshift range shown, corresponding to the redshift range in which the
parameter $F$ was measured (Lanzetta \etal\ 1991). The curves plotted
take no account of the possibility that the regions of star formation
fall off the slit.  The effect of the finite slit size is computed in
\S5.3. }
\label{fig:sfrcons}
\end{figure}

In Fig.~\ref{fig:sfrcons} we plot as solid curves
$\mathrm{\overline{SFR}}_{DLA}(LD)$ as a function of redshift for two
different cosmologies. The curves of $\dot{\Omega}_{g}(true)$ were taken
from fig.~1 ($q_{0}=0.5$) and fig.~2 ($q_{0}=0$) of Pei
\& Fall (1995), and are for a limiting cosmic density of neutral gas
at high redshift of $\Omega_{g\infty}=4\times10^{-3}~h^{-1}$. For the
hierarchical hypothesis the average SFR is lower by a factor $F=7.1$ for
$q_{0}=0.5$ and $F=3.8$
for $q_{0}=0$. These
predictions are plotted as dashed lines in Fig. 2, and are valid only
over the limited redshift range within which $F$ was measured.

Also plotted in Fig.~\ref{fig:sfrcons} are the upper limits to the SFR
for each DLA. In the case of absorbers observed with two slit positions
the limits are those for the region of overlap of the slit coverage,
i.e., the region of greatest sensitivity. We see that for $q_{0}=0$
our SFR limits lie well above the predicted mean SFR level, even for the
large-disk hypothesis. To reach the predicted SFRs appears to require an
8-metre class telescope.  For $q_{0}=0.5$ our limits also lie
above the prediction of the hierarchical hypothesis, but well below the
large-disk prediction. Therefore for this case (LD5) we could expect that
some of the DLAs observed would have SFRs brighter than our measured
limits. Nevertheless, whether the H$\alpha$ emission would have been
detected would depend on whether the region of SFR lay in the
slit. Under the hypothesis LD5 the typical radius of a DLA is very
large, $\sim 30~h^{-1}$ kpc. The area covered by the slit is
approximately $40\times 10~h^{-2}$~kpc$^{2}$. This has the consequence
that a high proportion of the regions of star formation will have fallen
off the slit, and the proportion is highest for the disks with the
largest SFRs. We quantify the incompleteness due to the finite solid
angle of the slit in the next sub-section.

\subsection{The effects of finite slit size on measured SFRs}

\begin{figure}
\epsfxsize=\columnwidth \epsfbox{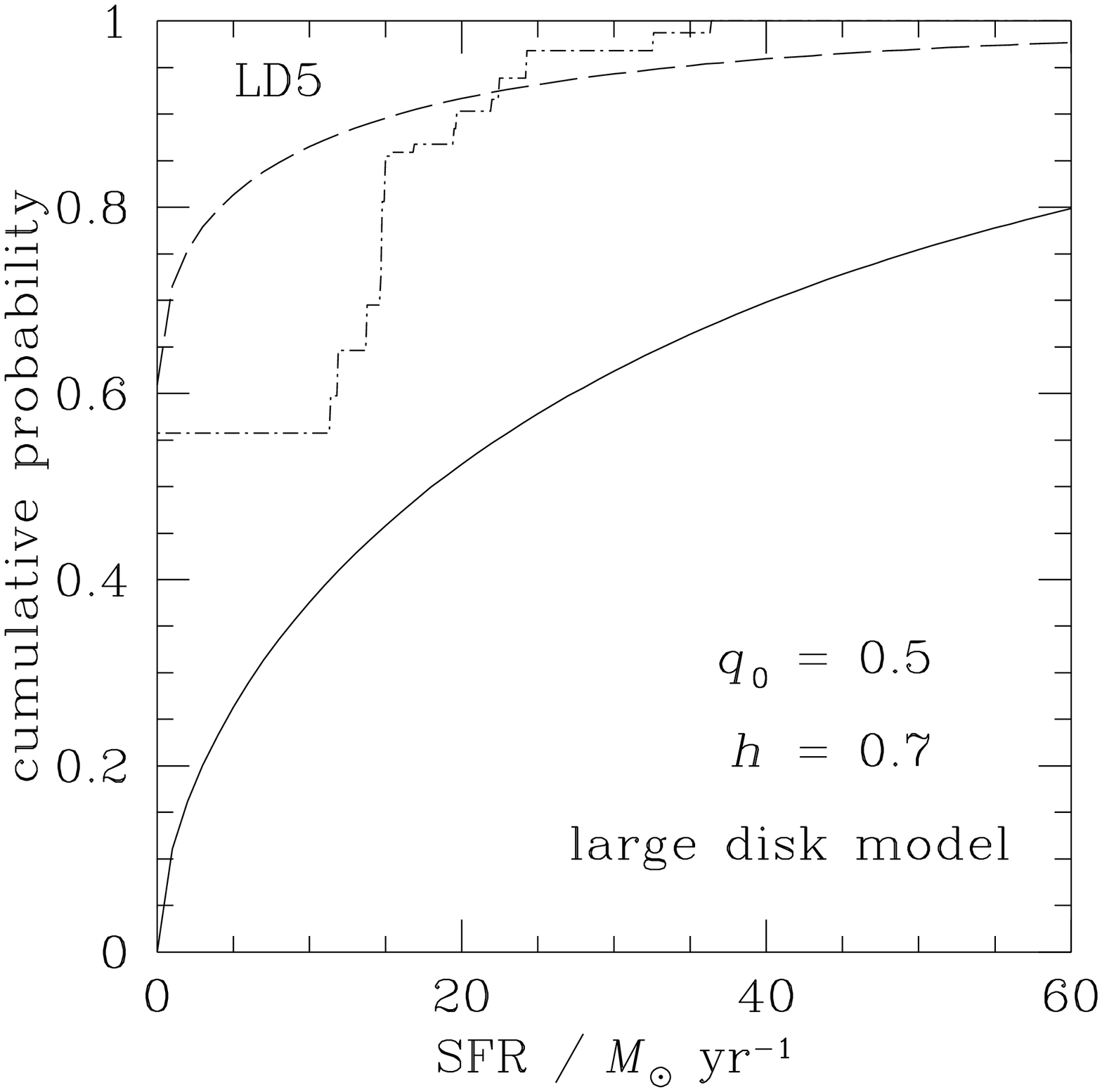}
\epsfxsize=\columnwidth \epsfbox{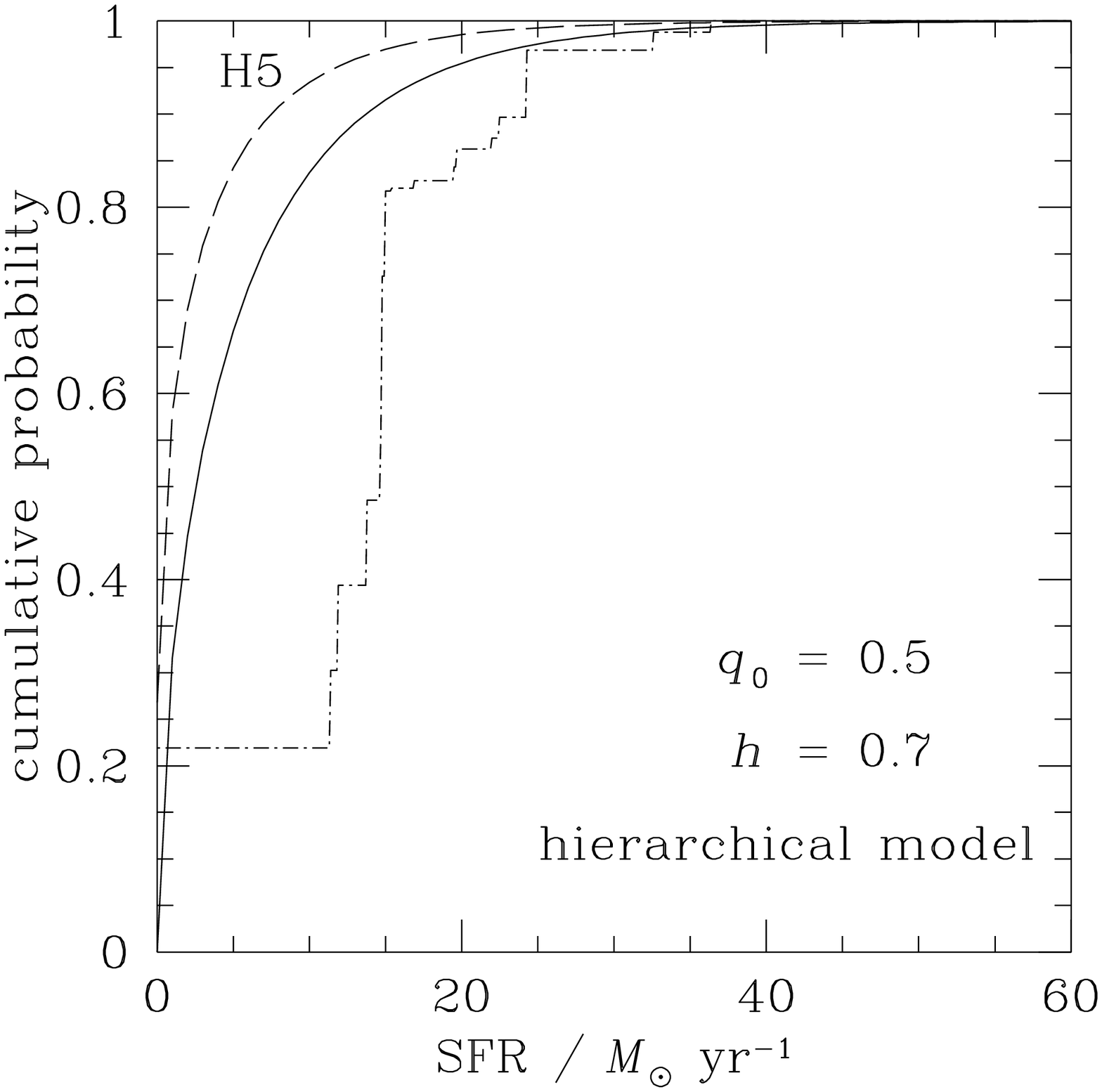}
\caption{Predicted and observed cumulative probability distributions of
SFRs for models LD5 (upper) and H5 (lower), for $h=0.7$ and
$q_{0}=0.5$. For each plot
the solid line shows the predicted distribution of SFRs for a sample of
DLAs at $z=2.3$. The dashed line shows the expected observed
distribution after accounting for the limited solid angle searched,
11~arcsec (spatial extent) x 2.5~arcsec (slit width). The stepped dot-dash
line plots the cumulative distribution of observational upper
limits. The value of 
the cumulative probability at which the stepped dot-dash line crosses the
dashed line is the expected proportion of DLAs that would not be
detected by our survey.}
\label{fig:sfcumprob5}
\end{figure}

\begin{figure}
\epsfxsize=\columnwidth \epsfbox{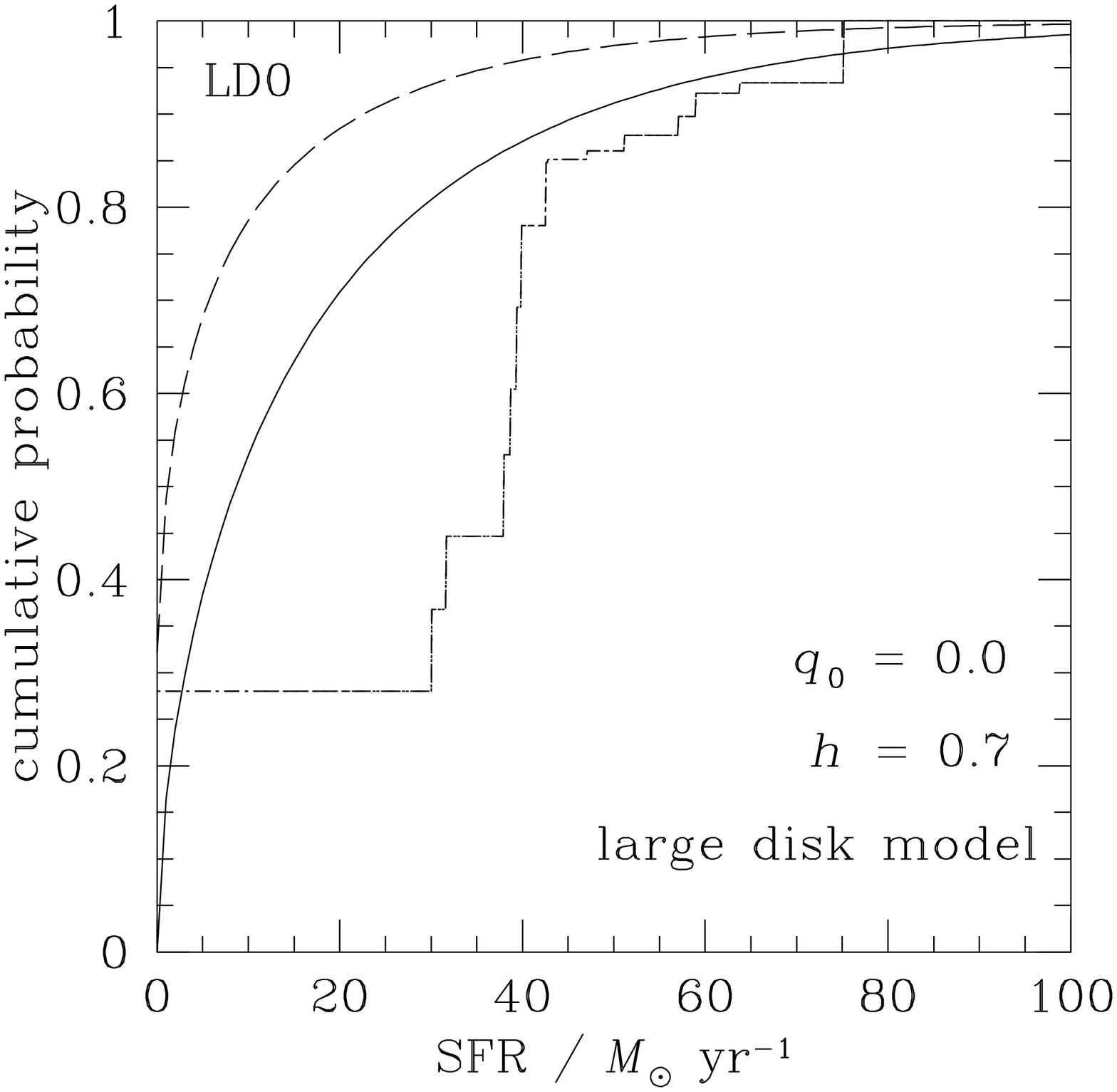}
\epsfxsize=\columnwidth \epsfbox{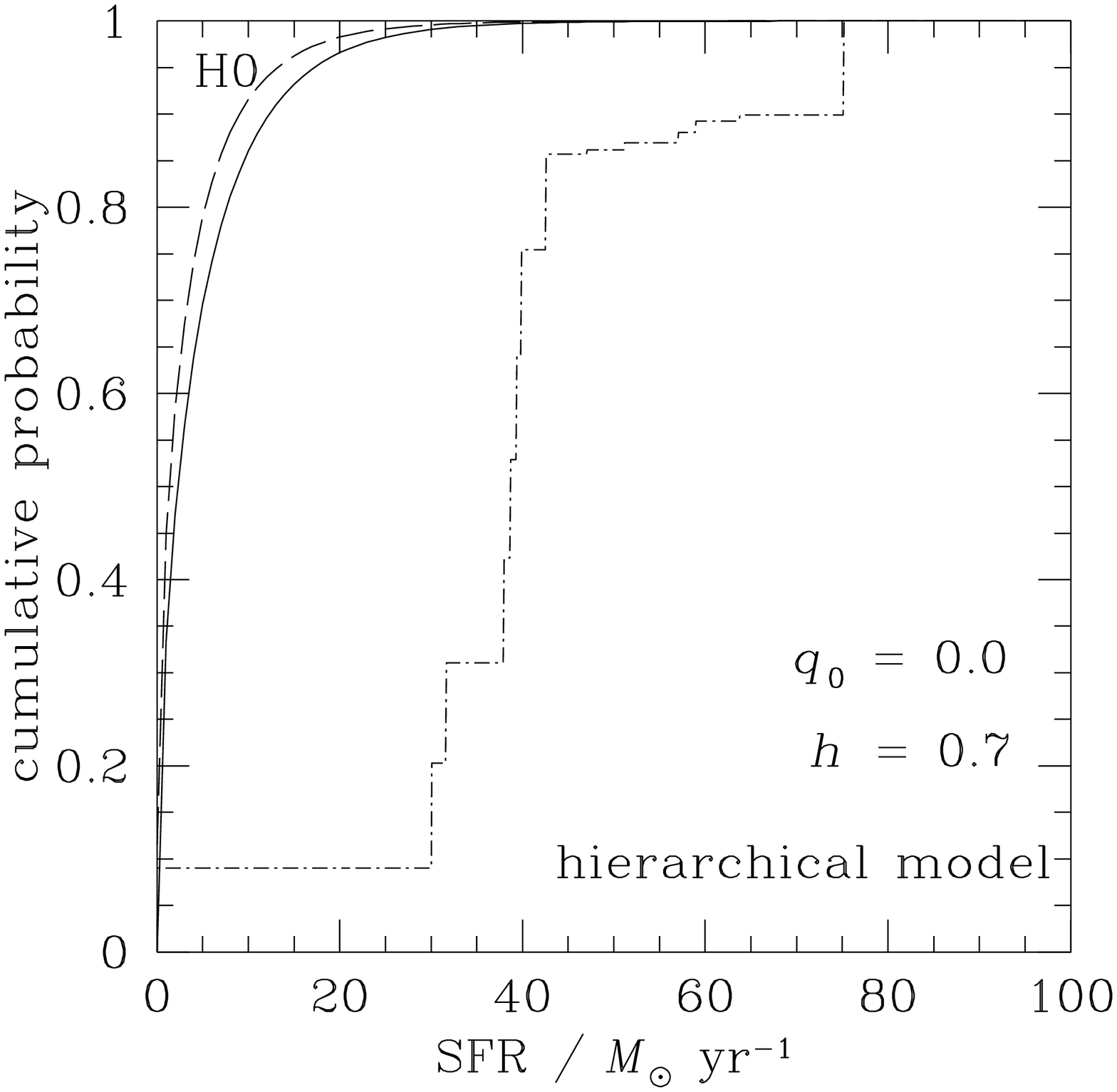}
\caption{Predicted and observed cumulative probability distributions of
SFRs for models LD0 (upper) and H0 (lower), for $h=0.7$ and
$q_{0}=0.0$. For each plot 
the solid line shows the predicted distribution of SFRs for a sample of
DLAs at $z=2.3$. The dashed line shows the expected observed
distribution after accounting for the limited solid angle searched,
11~arcsec (spatial extent) x 2.5~arcsec (slit width). The stepped
dot-dash line plots the cumulative distribution of observational upper
limits. The value of 
the cumulative probability at which the stepped line crosses the
dashed line is the expected proportion of DLAs that would not be
detected by our survey.}
\label{fig:sfcumprob0}
\end{figure}

Since the distributions of radius and SFR in the DLA population are
fully specified under either hypothesis, we can model the effects of the
finite solid angle searched. A Monte Carlo method is used. Firstly a
galaxy is selected at random from the radius distribution function
weighted by $R^2$ (to emulate the selection function from 1D quasar
sight-lines). The SFR for this size is computed, as appropriate for the
cosmology and the particular hypothesis, whether LD or H. In
Fig.~\ref{fig:sfcumprob5} the solid curves plot the cumulative
distributions of SFRs computed in this manner, for the mean redshift of
our sample $z=2.3$ and a cosmology of
$q_{0}=0.5$. Fig.~\ref{fig:sfcumprob0} repeats this for $q_{0}=0$. Note
that the H curves are simply the LD curves with the SFRs reduced by the
factor $F$.  

Next we need to establish whether the region of star formation would
fall on the slit. The galaxy is placed at a random inclination angle
and a random position angle. A point on the disk is selected at
random, and this is the sight line to the quasar. The slit is then
placed over this point. For the purposes of this calculation we assume
that all the star formation is in a compact region at the centre of
the DLA. If the centre of the DLA is now found to fall in the region
of the slit searched (or either slit if two PAs in Table~\ref{tab:obs}
were used) the SFR is recorded, otherwise the value zero. This then is
the predicted SFR that would have been measured by our survey if the
integration times had been extremely long, sufficient to detect the
faintest DLAs.  This distribution is plotted as the dashed curve in
each panel of Figs.~\ref{fig:sfcumprob5} \& \ref{fig:sfcumprob0}. Each
curve intersects the vertical axis at a finite value which is the
fraction of DLAs that fall off the slit (for example 56~per~cent in
the case of LD5). Note that the dashed curves, above this start point,
are skewed to lower SFRs than the solid curves. This is because it is
the DLAs with large SFRs that are more likely to be missed, because
larger SFRs occur in larger disks.

Finally we need to compare the depth actually reached against this
ideal, to quantify how many DLAs would have been detected under each
of the four hypotheses. This is achieved by comparing the SFR of each
hypothetical DLA against the sensitivity as follows.  For each
simulated DLA if the region of star formation falls within the section
of the slit searched the detection upper limit is recorded, otherwise
the value zero. In the case of DLAs observed at two PAs the SFR limit
corresponding to one or other slit, or to the region of overlap, is
recorded, as appropriate. The stepped dot-dash line records this
cumulative distribution of detection limits. Note that the dashed and
stepped dot-dash lines intersect the vertical axis at the same point.

The interpretation of these results is made by comparing the dashed
curve with the stepped line, for each plot. Since the stepped curve
records the sensitivity of the search, and the dashed curve plots the
predicted SFRs, where the stepped curve lies to the right of the
dashed curve, the survey limits were not sensitive enough to detect
the regions of star formation. Where the stepped curve lies to the
left of the dashed curve we would have expected to see an emission
line. For example for the hypothesis LD5 the two lines intersect near
a cumulative probability of 0.9. This implies that we would have
expected to detect H$\alpha$ emission from only $\sim$~10~per~cent of
our sample, or a single DLA. Our null result is only $1\sigma$
different from the LD expectation for $q_{0}=0.5$. The predicted
percentage of detections for the other three plots are even
smaller. The LD hypothesis is not ruled out for $q_{0}=0$, and we
conclude that our survey is insufficiently sensitive to test
hypothesis H for either cosmology.

\section{Summary and conclusions}
\label{sec:conclusions}

To summarise, we have searched for H$\alpha$ emission from eight damped
Ly$\alpha$ absorbers at $z>2$.  No H$\alpha$ emission was detected above
$3\sigma$ limits in the range $(6.5 - 16)\times 10^{-20}$~W~m$^{-2}$. We
have compared our results against the predictions of the models of Pei
\& Fall (1995) under the `large-disk' (LD) and `hierarchical' (H)
hypotheses.  Compared to the previous most sensitive spectroscopic
search, our sample is twice as large, our limits are a factor greater
than two deeper, and the solid angle surveyed is over three times as
great. Despite this our results are not in conflict with either
hypothesis. In the case of LD this is because the solid angle searched,
$2.5 \times 11$~arcsec$^{2}$, is too small. In the case of H, although
most of the regions of star formation fall in the slit (the solid and
dashed curves, Figs.~\ref{fig:sfcumprob5} \& \ref{fig:sfcumprob0}, are
much closer), the sensitivity is insufficient.

The plots of Fig. 3 demonstrate in a quantitative way how inefficient
spectroscopy is for testing the large-disk hypothesis. The spectroscopic
searches for Ly$\alpha$ emission from DLAs by Pettini \etal\ (1995)
and Lowenthal \etal\ (1995) are even less efficient for this purpose,
because the solid angle searched was an order of magnitude smaller than
for our H$\alpha$ work. Despite reaching less sensitive flux limits
narrow-band H$\alpha$ imaging searches are probably more efficient for
testing the large disk hypothesis because the brightest DLA counterparts
are expected to have the largest impact parameters. We will address this
in a future paper where we intend to incorporate the results of the
existing narrow-band searches (e.g., Bunker \etal\ 1995; Bunker 1996;
Bechtold \etal\ 1998; Mannucci \etal\ 1998; Teplitz, Malkan \& McLean
1998) into the Monte-Carlo analysis.

The measured impact parameters of the few DLAs that have been detected
in imaging searches provide a further test of the large-disk
hypothesis. M\o ller \& Warren (1998) showed that the small measured
impact parameters of the few confirmed or candidate counterparts of
high-redshift DLAs implied that for $q_{0}=0.5$ the space density of
DLAs is a factor greater than five times the space density of spiral
galaxies today. This result is in conflict with LD5. If the
large-disk hypothesis is eventually ruled out for a low-density
cosmology as well, the plots of Figs.~\ref{fig:sfcumprob5} \&
\ref{fig:sfcumprob0} indicate that deep spectroscopy may be
the most efficient method to test the hierarchical hypothesis: because
in this case the DLAs are small and will usually be covered by the
slit, and because spectroscopy can reach fainter flux limits than
narrow-band imaging. Flux limits a factor four deeper than those
reached here would be required.

Perhaps the most promising line of attack is imaging with the HST
cameras NICMOS and STIS. Here the high spatial resolution and
sensitivity allow the detection of very faint sources $m_H=23.5$~mag,
$m_V=28$~mag as close as 0.4~arcsec from the line of sight to the quasar.
In this way candidate counterparts to the DLA would be identified
allowing a narrower slit for the subsequent spectroscopy. By this means
it would be possible to reach the required faint flux limits with an
8-metre class telescope.

\appendix
\section{Near-Infrared Spectroscopy of the Quasars}
\label{app:qsospec}

\begin{figure}
\epsfxsize=\columnwidth \epsfbox{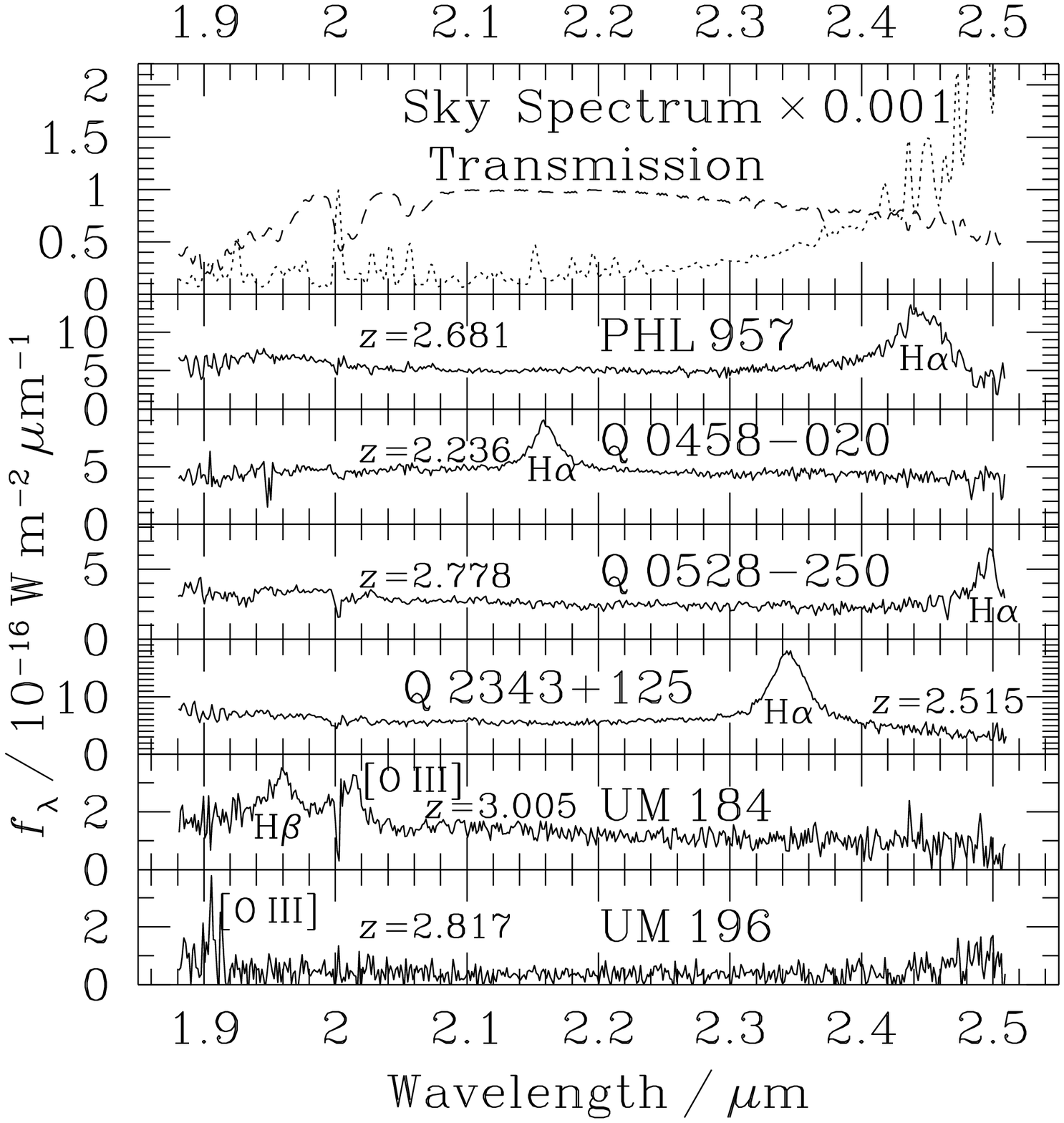}
   \caption{$K$-band spectra of the six target quasars, in units of
   $f_{\lambda}$ with the prominent
   emission lines labelled. The top panel shows the sky spectrum (dashed
   line, scaled down by a factor of 1000) and the atmospheric
   transmission (dotted line).}
   \label{fig:dlaspec}
\end{figure}

Optimally extracted near-infrared spectra of six $z>2$ quasars were
obtained (Figure~\ref{fig:dlaspec}), enabling the continua to be
measured (Table~\ref{tab:qsodata}). Strong, broad H$\alpha$ was detected
from PHL957, 0458$-$020, 0528$-$250 and 2343$+$125. The
[O{\scriptsize~III}]~$\lambda$~500.7~nm line was detected in PHL957
and UM196, as well as UM184 which also exhibited broad
H$\beta$. Table~\ref{tab:qsoemlines} gives the equivalent widths,
fluxes, velocity widths and wavelengths for these lines. The redshifts
inferred are in many instances discrepant from the literature values,
which are primarily derived from rest-frame UV lines such as Ly$\alpha$,
detected in the optical.
 
A case of particular interest is 0528$-$250: this sight-line intercepts
a second damped absorber, a $z_{\rm QSO}\approx z_{\rm DLA}$ system
(M\o ller \& Warren 1993) as well as the DLA at $z_{\rm abs}=2.14$
which was the target of the near-infrared spectroscopy. The redshift
of the second absorber is $z=2.811$, and a puzzle has been that this
is redward of the literature value for the quasar, $z=2.77$ (Morton
\etal\ 1980). This was determined from the high-ionization
C{\scriptsize~IV}~$\lambda$~154.9~nm line, which is typically
blue-shifted with respect to the systemic redshift. The Ly$\alpha$
emission line is unusable, being almost completely absorbed by the
damped system. From the near-infrared spectroscopy, the redshift is
re-measured from the H$\alpha$ line to be $z\approx 2.806$, and the
[O{\scriptsize~III}]~$\lambda$~500.7~nm line has a redshift of
$z=2.788$.  The real QSO redshift is closer to that of the damped
system than formerly thought.

In all six QSOs, the low-ionization lines
[O{\scriptsize~I}]~$\lambda$~630.0~nm and
[S{\scriptsize~II}]~$\lambda\lambda$~671.6/673.1~nm were undetected,
with the $3\sigma$ upper limit for [O{\scriptsize~I}] typically
$1.8\times 10^{-19}$~W~m$^{-2}$ if unresolved. The [S{\scriptsize~II}]
doublet was generally too near the broad-wings of H$\alpha$ for upper
limits to be placed with confidence.

\begin{table*}
\begin{tabular}{||l|l|l|l|c|c|c|c|c|c||}
\multicolumn{10}{c}{} \\
\hline\hline
& & & & & & \multicolumn{2}{c|}{} & \multicolumn{2}{c||}{} \\
QSO & $K$-band & $K$-band &
Interval & $f_{\lambda}~/~10^{-16}$ & $f_{\nu}$ 
& \multicolumn{2}{c|}{SNR~/~$\sigma$}
& \multicolumn{2}{c||}{$L_{\nu}~/~10^{24}$~W~Hz$^{-1}$} \\
& Spec.\ /~mag & Imaging~/~mag & /~$\mu$m & W~m$^{-2}$~$\mu$m$^{-1}$ &
/~$\mu$Jy & pix$^{-1}$ & res$^{-1}$ & ($q_{0}=0.5$) & ($q_{0}=0$) \\
\hline\hline
PHL957 & $14.70$ ($300~\sigma$) &
$14.74$ ($300~\sigma$)\,$^a$ & 
$2.10\rightarrow 2.30$ & 5.0 & 800 & 22 & 42 & 2.91 & 9.22 \\
& $14.65$ ($K$) & & & & & & & & \\
0458$-$020 & $14.73$ ($650~\sigma$)
& $14.74$
($140~\sigma$)\,$^b$ & $2.22\rightarrow 2.40$ &
4.5 & 790 & 19 & 36 & 2.17 & 6.00 \\
& $14.78$ ($K$) & & & & & & & & \\
0528$-$250 & $15.46$ ($350~\sigma$) &
$15.46$ ($200~\sigma$)\,$^a$ &
$2.05\rightarrow 2.30$ &2.5 & 400 & 12 & 23 & 1.53 & 6.72 \\
2343$+$125 & $14.42$ ($500~\sigma$) & 
-- 
& $2.05\rightarrow 2.20$ & 5.6 & 840 & 18 & 34 & 2.77 & 8.30 \\
UM184 & $16.24$ ($50~\sigma$) &
$16.16$ ($75~\sigma$)\,$^c$ & 
$2.07\rightarrow 2.30$ & 1.3 & 200 & 5.5 & 10 & 0.86 & 3.03 \\
UM196 & $17.59$ ($35~\sigma$)$^{d}$ &
$16.75$ ($40~\sigma$)\,$^c$ &
$2.05\rightarrow 2.35$ & 0.36 & 58 & 1.7 & 3.2 & 0.23 & 0.76 \\
& $17.59$ ($K$)$^{d}$ & & & & & & & & \\
\hline\hline
\multicolumn{10}{c}{} \\
\end{tabular}
 
$^a$ ESO~2.2-m~/~IRAC2B 
November 1993, $K'$ filter
\hspace{2cm} $^b$  ESO~2.2-m~/~IRAC2B 
November 1994, $K'$ filter
\newline $^c$ UKIRT~/~IRCAM3 \hspace{2cm} $^{d}$ Possibly affected by cloud
July 1995, $K$ filter
\caption{Measured properties of the background QSOs from the $K$-band
spectroscopy. The signal-to-noise is quoted both per pixel and per
resolution element ($\approx 650~{\rm km~s}^{-1}$, $\approx
4$~pix). The ESO~2.2-m~/~IRAC2B broad-band imaging uses the
$K^{\prime}$ filter, and the flux over the same wavelength range is
measured from the near-infrared spectra for comparison purposes. The
magnitude from the spectra integrated over the entire $K$-window is
also quoted. The only major discrepancy between the broad-band
magnitudes from the imaging and the spectroscopic fluxes is for UM196.}
\label{tab:qsodata}
\end{table*}

\begin{table*}
\begin{center}
\begin{tabular}{||l|l|l|c|c|c|c|c|c|c||}
\multicolumn{10}{c}{} \\
\hline\hline
& & & & & & \multicolumn{2}{c|}{} & & \\
QSO & Redshift & Wavelength & Line ID & FWHM & flux &
\multicolumn{2}{c|}{Luminosity} & Redshift
& EW$_{\rm rest}$ \\
& (literature) & /~$\mu$m & & /~km~s$^{-1}$ & $f_{\rm
line}$ &  
\multicolumn{2}{c|}{$L_{\rm line}~/~10^{35}$~W} & (measured) &
 \\
& & & & & /~$10^{-20}$~W~m$^{-2}$ & ($q_{0}=0.5$) & ($q_{0}=0.1$) &
& (nm) \\ 
\hline\hline
PHL957 & 2.681 & 2.443 & H$\alpha$ & 5500 & 3400 & 455 & 1440 & 2.723 & 18.5 \\
 & & 1.837 & [O{\scriptsize~III}] & $<$~650 & 110 & 15 & 48 & 2.669 & 0.2 \\
0458$-$020 & 2.286 & 2.1595 & H$\alpha$ & 3750 & 1040 & 97 & 270 & 2.291 &
8.2 \\ 
0528$-$250 & 2.77 & 2.498 & H$\alpha$ & 2300 &
740 & 107 & 354 & 2.806 & 9.2 \\ 
    & & 1.897 & [O{\scriptsize~III}] & $<$~650 & 40 & 5.8 & 19 & 2.788 & 1.1 \\
2343$+$125 & 2.515 & 2.344 & H$\alpha$ & 4600 & 4400 & 510 & 1530 & 2.517 &
21.4 \\
UM184 & 3.005 & $\ga$~1.96\,$^{\dagger}$ & H$\alpha$ & -- & -- &
-- & -- &
$\ga$~2.99 & -- \\
UM196 & 2.817 & $\ga$~1.906\,$^{\dagger}$ & [O{\scriptsize~III}] & -- &
$>$~50 & $>$~7.5  & $>$~25 & $>$~2.807 &
$>$~13.2 \\ 
\hline\hline
\multicolumn{10}{c}{} \\
\end{tabular}
 
$^{\dagger}$\,Complicated by presence of atmospheric absorption
feature at $\simeq$~1.98~$\mu$m.
\end{center}
\caption{Measured properties of the emission lines of the background
QSOs in the near-infrared CGS\,4 spectroscopy with UKIRT
(October 1995). Tabulated are the measured wavelengths, widths and
fluxes for the lines, and the corresponding line luminosity, redshift
and rest-frame equivalent width ($EW_{\rm rest}$) for the line
identifications.}
\label{tab:qsoemlines}
\end{table*}

\section{The Companion Galaxy to the PHL957 Damped Absorber}
\label{sec:phl957c1spec}
 
\begin{figure}
\epsfxsize=\columnwidth \epsfbox{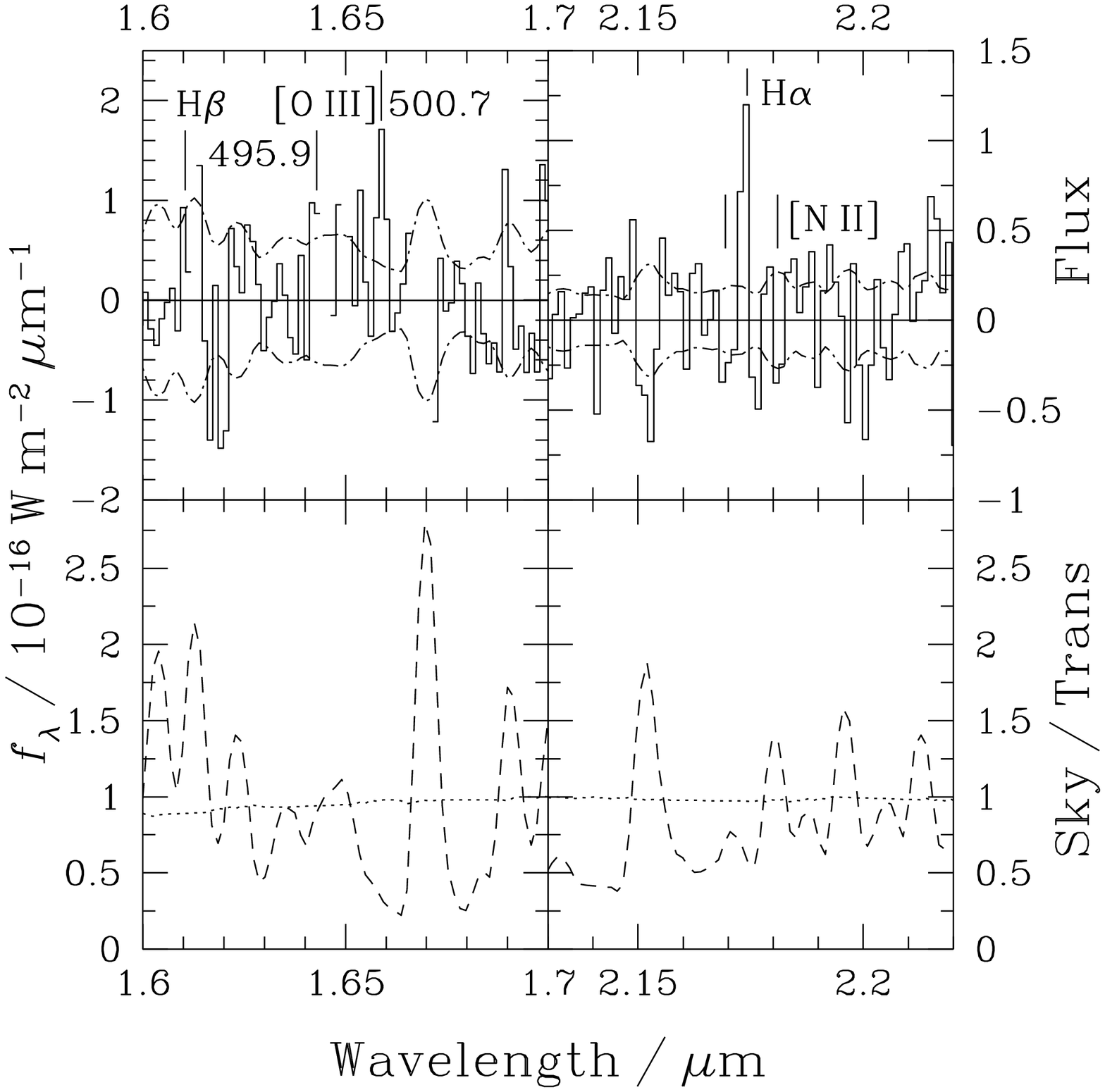}
   \caption{The H$\alpha$ (top-right panel)
and [O{\scriptsize~III}]~$\lambda$~500.7~nm
emission (top-left panel) from the $z=2.31$ companion galaxy, C1, of the
   DLA PHL957. The dot-dash lines are the $\pm 1\sigma$ noise per pixel, and
   the lower panels show the fractional atmospheric transmission (dotted
   line) and the sky spectrum scaled down by a factor of 1000 (short-dash
   line).}
   \label{fig:phl957oiiiandc1}
\end{figure}

By appropriate slit orientation, a spectrum of the companion C1
($z=2.313$, Lowenthal \etal\ 1991) was obtained simultaneously with a
spectrum of PHL957 (Figure~\ref{fig:phl957oiiiandc1}). The large impact
parameter of 48~arcsec -- a projected physical distance of
$190~h^{-1}$~kpc ($320~h^{-1}$~kpc) for $q_{0}=0.5$ ($q_{0}=0.0$) --
argues against this being the
actual galaxy responsible for the absorption in the QSO spectrum at
$z=2.309$, but rather a galaxy which is in the same group as the DLA.
 
The H$\alpha$ line detected by Hu \etal\ (1993) and Bunker \etal\ (1995)
is confirmed at $3\sigma$, and is spectrally unresolved
($<650$~km~s$^{-1}$), centred at 2.1735~$\mu$m. The inferred redshift of
$z=2.312$ agrees to within $\approx 100$~km~s$^{-1}$ of the measurement
from Ly$\alpha$ of $z=2.3128$ (Lowenthal \etal\ 1991), with the
difference probably attributable to absorption of the blue wing of
Ly$\alpha$. The integrated flux is $(2.3\pm 0.8)\times
10^{-19}$~W~m$^{-2}$, consistent with the estimates of $(3.6\pm
1.5)\times 10^{-19}$~W~m$^{-2}$ from Hu \etal\ (1993) and $(2.1\pm
0.6)\times 10^{-19}$~W~m$^{-2}$ from Bunker \etal\ (1995). The
[N{\scriptsize~II}] doublet is undetected in our near-infrared
spectroscopy. The H$\alpha$ line flux corresponds to a star formation
rate of $20\pm 7~h^{-2}~{\rm M_{\odot}~yr}^{-1}$ ($55\pm 19~h^{-2}~{\rm
M_{\odot}~yr}^{-1}$) using the Kennicutt (1983) relation. However, the
presence of moderately strong C{\scriptsize IV}~$\lambda$~154.9~nm and
He{\scriptsize II}~$\lambda$~164.0~nm emission, and the relatively large
rest-frame equivalent width of 14~nm for Ly$\alpha$ (Lowenthal \etal\
1991) argue for a significant AGN contribution.

There is a marginal ($3\sigma$) detection of unresolved
[O{\scriptsize~III}]~$\lambda$~500.7~nm at $\lambda= 1.6589~\mu$m with
$f({\rm [O{\scriptstyle~III}]~500.7})=(4.3\pm 1.5)\times
10^{-19}$~W~m$^{-2}$, while the 495.9~nm line of this
[O{\scriptsize~III}] doublet and H$\beta$ are undetected with a
$3\sigma$ upper limit of $4.4\times 10^{-19}$~W~m$^{-2}$. Therefore,
$f$[O{\scriptsize~III}]~/~$f$(H$\alpha$)~$= 2.5^{+2}_{-1}$.  Although the
continuum of galaxy C1 is undetected in the near-infrared spectroscopy,
a marginal detection of the continuum was made in the broad-band imaging
of Bunker \etal\ (1995). The $K^{\prime}$ magnitude is $21.0^{m}$
($2\sigma$), so with the line contribution removed the $3\sigma$
upper-limit on the continuum flux is $f_{\lambda}~<~1.8\times
10^{-18}~{\rm W~m}^{-2}~\mu{\rm m}^{-1}$, $f_{\nu}~<~3.1~\mu$Jy.  This
would imply the continuum luminosity density is $L_{\nu}(3\sigma)<8.2
\times 10^{21}~h^{-2}$~W~Hz$^{-1}$ ($L_{\nu}~<~22\times
10^{21}~h^{-2}$~W~Hz$^{-1}$). The $2\sigma$ lower-limit on the
rest-frame equivalent width of H$\alpha$ is 22~nm (Bunker \etal\ 1995).

\subsection*{ACKNOWLEDGMENTS}

We wish to thank Tom Geballe, Gillian Wright, Thor Wold and John
Davies for their help with the observations. The UKIRT is operated by
the Royal Observatories on behalf of the UK Particle Physics and
Astronomy Research Council (PPARC). We thank Richard McMahon for
making the co-ordinates of quasar 2343$+$125 available to us. We have
benefited from discussions with Mike Fall, Max Pettini and Hyron
Spinrad. We thank the referee, Palle M\o ller, for his constructive
comments on the manuscript. AJB
gratefully acknowledges the award of a PPARC PhD studentship while at
Oxford, and subsequently a NICMOS postdoctoral fellowship at UC Berkeley.

\bsp


\begin{thebibliography}{}

\bibitem[\protect\citename{Bechtold \etal\ }1998]{be98}Bechtold J.,
Elston R., Yee H. K. C., Ellingson E., Cutri R. C., 1998, in D'Odorico
S., Fontana A., Giallongo E., eds., ASP Conf.\ Ser.\ Vol.\ 146, The
Young Universe: Galaxy Formation and Evolution at Intermediate and
High Redshift, Astron.\ Soc.\ Pac.., San Francisco, p.~241

\bibitem[\protect\citename{Bruzual \& Charlot }1993]{br93}A.\ Bruzual G.,
 Charlot S., 1993, ApJ, 405, 538

\bibitem[\protect\citename{Bunker \etal\ }1995]{bu95}Bunker A.~J.,
Warren S.~J., Hewett P.~C., Clements D.~L., 1995, MNRAS, 273, 513

\bibitem[\protect\citename{Bunker }1996]{bu96}Bunker A.~J., 1996,
D.Phil.\ Thesis, University of Oxford, ``Searches for Distant
Galaxies'', abstract in 1997, PASP, 109, 628. WWW address {\tt
http://astro.berkeley.edu/\~\,bunker/thesis.html}

\bibitem[\protect\citename{Calzetti }1997]{ca97}Calzetti D., 1997, AJ,
113, 162

\bibitem[\protect\citename{Charlot \& Fall }1991]{ch91}Charlot S.,
Fall S.~M., 1991, ApJ, 378, 471

\bibitem[\protect\citename{Djorgovski \etal\ }1996]{dj96}Djorgovski
S.~G., Pahre M.~A., Bechtold J., Elston R., Nature, 382, 234

\bibitem[\protect\citename{Elston \etal\ }1991]{el91}Elston R.~J.,
Bechtold J., Lowenthal J.~D., Rieke M., 1991, ApJ, 373, L39

\bibitem[\protect\citename{Gallego \etal\ }1995]{ga95}Gallego J.,
Zamorano J., Arag\'{o}n-Salamenca A., Rego M., 1995, ApJ, 455, L1

\bibitem[1996]{gi96} Giavalisco M., Steidel C.~C., Macchetto F.~D.,
   1996, ApJ 470, 189

\bibitem[\protect\citename{Haehnelt, Steinmetz \& Rauch
}1998]{ha98}Haehnelt M.~G., Steinmetz M., Rauch M., 1998, ApJ, 495, 647

\bibitem[\protect\citename{Hunstead, Pettini \& Fletcher
}1990]{hu90}Hunstead R.~W., Pettini M., Fletcher A.~B., 1990, ApJ, 356,
23

\bibitem[\protect\citename{Hu \etal\ }1993]{hu93}Hu E.~M., Songaila
A., Cowie L.~L., Hodapp K-W., 1993, ApJ, 419, L13

\bibitem[\protect\citename{Kennicutt }1983]{ke83}Kennicutt R.~C.,
1983, ApJ, 272, 54

\bibitem[\protect\citename{Lanzetta \etal\ }1991]{la91}Lanzetta K.~M.,
Wolfe A.~M., Turnshek, D.~A., Lu L., McMahon R.~G., Hazard C., 1991,
ApJS, 77, 1

\bibitem[\protect\citename{Lanzetta, Wolfe \& Turnshek
}1995]{la95}Lanzetta K.~M., Wolfe A.~M., Turnshek D.~A., 1995, ApJ,
440, 435

\bibitem[\protect\citename{Lilly \etal\ }1996]{li96}Lilly S.~J.,
Le~F\`{e}vre O., Hammer F., Crampton D., 1996, ApJ, 460, L1 

\bibitem[\protect\citename{Lowenthal \etal\ }1991]{lo91}Lowenthal J.~D.,
Hogan C.~J., Green R.~F., Caulet A., Woodgate B.~e., Brown 
L., Foltz C.~B., 1991, ApJ, 377, L73

\bibitem[\protect\citename{Lowenthal \etal\ }1995]{lo95}Lowenthal J.~D.,
Hogan C.~J., Green R.~F., Woodgate B.~E., Caulet A., Brown 
L., Bechtold J., 1995, ApJ, 451, 484

\bibitem[\protect\citename{Madau \etal\ }1996]{ma96}Madau P., Ferguson
H.~C., Dickinson M.~E., Giavalisco M., Steidel C.~C., Fruchter A., 1996,
MNRAS, 283, 1388

\bibitem[\protect\citename{Mannucci \etal\ }1998]{ma98}Mannucci F.,
Thompson D., Beckwith S.~V.~W., Williger G.~M., 1998, ApJ, 501, L11

\bibitem[\protect\citename{Meyer \& Roth }1990]{mey90}Meyer D.~M.,
Roth  K.~C., 1990, ApJ, 363, 57

\bibitem[\protect\citename{M\o ller \& Warren }1993]{mo93}M\o ller P.,
Warren S.~J., 1993, A\&A, 270, 43

\bibitem[\protect\citename{M\o ller \& Warren }1998]{mo98}M\o ller P.,
Warren S.~J., 1998, MNRAS, 299, 661

\bibitem[\protect\citename{Morton \etal\ }1990]{mo90}Morton D.~C., Chen
J., Wright A.~E., Peterson B.~A., Jauncey D.~L., 1980, MNRAS, 193, 399

\bibitem[\protect\citename{Pei \& Fall }1995]{pei95}Pei Y.~C., Fall
S.~M., 1995, ApJ, 454, 69

\bibitem[\protect\citename{Pettini, Boksenberg \& Hunstead
}1990]{pe90}Pettini M., Boksenberg A., Hunstead R.~W., 1990, ApJ, 
356, 23

\bibitem[\protect\citename{Pettini \etal\ }1994]{pe94}Pettini M.,
Smith L.~J., Hunstead R.~W., King D.~L., 1994, ApJ, 426, 79

\bibitem[\protect\citename{Pettini \etal\ }1995]{pe95}Pettini M.,
Hunstead R.~W., King D.~L., Smith L.~J., 1995, in Meylan G., ed., ESO
Astrophysics Symposia Series, QSO Absorption Lines, Springer-Verlag,
Berlin, p.~55

\bibitem[\protect\citename{Pettini \etal\ }1997]{pe97}Pettini M.,
King D.~L., Smith L.~J., Hunstead R.~W., 1997, ApJ, 478, 536

\bibitem[\protect\citename{Prochaska \& Wolfe}1997]{pr97}Prochaska
J.~X., Wolfe A.~M., 1997, ApJ, 487, 73

\bibitem[\protect\citename{Salpeter }1955]{sa55}Salpeter E.~E., 1955,
ApJ, 121, 161

\bibitem[\protect\citename{Sargent, Boksenberg \& Steidel
}1988]{sa88}Sargent W.~L.~W., Boksenberg A., Steidel C.~C., 
1988, ApJS, 68, 539

\bibitem[\protect\citename{Sargent, Steidel \&
Boksenberg}1989]{sa89}Sargent W.~L.~W., Steidel C.~C., Boksenberg A.,
1989, ApJS, 69, 703

\bibitem[\protect\citename{Scalo }1989]{sc86}Scalo J.~M., 1986, Fund.\
Cosmic Phys., 11, 1

\bibitem[\protect\citename{Smith \etal\ }1989]{sm89}Smith H.~E., Cohen
R.~D., Burns J.~E., Moore D.~J., Uchida B.~A., 1989, ApJ, 347, 87

\bibitem[\protect\citename{Steidel, Pettini \& Hamilton
}1995]{st95}Steidel C.~C. Pettini M., Hamilton D., 1995,
AJ, 110, 2519

\bibitem[\protect\citename{Steidel \etal\ }1996]{st96}Steidel C.~C.,
Giavalisco M., Pettini M., Dickinson M.~E., Adelberger K.~L., 1996,
ApJ, 462, L17

\bibitem[\protect\citename{Steidel \etal\ }1999]{st99}Steidel C.~C.,
Adelberger K.~L., Giavalisco M., Dickinson M.~E., Pettini M., 1999,
ApJ, {\em in press}

\bibitem[\protect\citename{Teplitz \etal\ }1998]{te98}Teplitz H.~I.,
Malkan M.~A., McLean I.~S., 1998, ApJ, 506, 519

\bibitem[\protect\citename{Turnshek \etal\ }1989]{tu89}Turnshek
D.~A., Wolfe A.~M., Lanzetta K.~M., Briggs F.~H., Cohen R.~D., Foltz
C.~B., Smith H.~E., Wilkes B.~J., 1989, ApJ, 344, 567


\bibitem[\protect\citename{Warren \& M\o ller }1996]{wa96}Warren
S.~J., M\o ller P., 1996, A\&A, 331, 25

\bibitem[\protect\citename{Wolfe \etal\ }1986]{wo86}Wolfe A.~M.,
Turnshek D.~A., Smith H.~E., Cohen R.~D., 1986, ApJS, 61, 249

\bibitem[\protect\citename{Wolfe \etal\ }1992]{wo92}Wolfe A.~M.,
Turnshek D.~A., Lanzetta K.~M., Oke J.~B., 1992, ApJ, 385, 151

\bibitem[\protect\citename{Wolfe \etal\ }1993]{wo93}Wolfe A.~M.,
Turnshek D.~A., Lanzetta K.~M., Lu L., 1993, ApJ, 404, 480

\end{thebibliography}
\end{document}